# On Portrait of a Specialist in Open Data

Irina Radchenko[1[0000-0001-8658-4083]], Anna Koroleva[1[0000-0002-1673-2308]]
and Yaroslav Baranov[1[0000-0002-7328-7300]]

[1] ITMO University, Kronverksky 49, St.Petersburg, Russia
iradche@gmail.com, anna.koroleva@gmail.com, kciray8@gmail.com

**Abstract.** The article is written to identify the requirements for Open Data Specialist. The ability to use and work with open data affects many areas: sociology, urban studies, geography, statistics, public administration, data journalism, etc. It is especially important to develop and implement training courses on open data for non-IT students. Typically, the specialization of these students contains insufficient number of lessons on working with data and with open data. Students (hereinafter - researchers) feel a great need to eliminate "digital illiteracy" and to master the skills of working with open data. The development of a specialty on open data is designed to solve the problem of lack of knowledge and skills in working with open data.

In this paper, the authors attempt to generalize the requirements for an expert in open data and offer an overview of information sources on the topic of hiring such specialists.

The authors justify the need to create a specialty on open data for non-core students as well. It is supposed that the specialty will be read in English, a non-native language for students.

## 1 Introduction

Open Data, according to "DBpedia: A Nucleus for a Web of Open Data" is the idea that some data should be freely available to everyone to use and republish as they wish, without restrictions from copyright, patents or other mechanisms of control [1].

Being an idea Open Data is paid a lot of attention nowadays as it plays a great role. Open data is a huge resource, especially open government data. Thus, a big number of companies collects, processes and stores different types of data that should be managed.

One of the key features that should be taken into account is a legality. So, government service workers need to be educated and know how to use open data as it should be transparent and accessible. However, employees as well as students feel lack of knowledge. Typically, the specialization of these people contains insufficient practice and experience on working with data and with open data. Feeling a great need to eliminate "digital illiteracy" and to master the skills of working with open data it is especially important to develop and implement training courses on open data for non-IT specialists.

Open Data includes different areas of life: transport, geodata, culture, science, financial, statistics, weather, environment, education, business and so on. Moreover, all these spheres and many more affect Open Data as a valuable resource to gain awareness and promotion. So, every field without reference to a particular activity requires a skillful expert whose job is to analyze, process, manage and control data and open data especially as this type of data provides public with free information and products.

One of the matter we pay attention in this paper is teaching and using open data as a resource to develop specialists. As nowadays in an attempt to be an up to date specialist people try to search information using both native language and English, data is presented in one the chapters. Companies give preference to young and experienced experts who have the language and professional competences. Although it is a big dilemma to teach a subject in English, as it is an international language, or to teach English in a professional area (ESP). Anyway, in our rapidly growing world it is almost impossible not to use online resources using open data that provides us with current and topical information. Our study is built on work with students from IT and CS departments university ITMO.

This paper is a base for our further study on open data implementation in education using non-mother tongue.

## 2    A Case Study: Open Data Specialist Survey

### 2.1    Data Collection and Design

The research is designed to identify the skills and competences a specialist, in IT and non-IT areas, should possess in order to fulfil working tasks efficiently. We organized a questionnaire which includes 12 questions specified into three groups: demographic, professional competences of open data specialist (ODS) and English language awareness. We used the most common google forms and designed the questions as multiple choice, dichotomous and open-ended questions. The detailed results are presented in chats with descriptions in the section 2.2.

It is aimed to inquire what core competences are in demand, and verify a set of skills to work with open data. We placed the questionnaire in various publics such as:

- Telegram: OpenDataRussia Chat (330 members), Data Science Chat (1077 members), DDJ (63 members), datajouru (108 members), SPbLUG_chat (89 members), RU.SYSADMIN (728 members);
- Slack: opendatascience #career (1875 members), #interesting_links (4887 members);
- Facebook: distant education professionals group (8315 members), Open Data School page (532 likes) where we used Facebook promotion during 4 days for 1 168 people;
- Private contacts.

The approximate number of audience is 19 172 people, whereas the number of respondents is 75 due to a Russian questionnaire and 21 due to its English version in total.

Subjects matter of the channels are the following:

- Open Data,
- Data Science,
- Data Journalism,
- Community Open Data Science,
- eLearning,
- Open Source,
- IT.

The respondents are mainly representatives of IT and CS spheres.

This type of specialists attracts our particular attention, as a part of the further study related to methodology of teaching ESP (English for Specific Purposes) and EMI (English Medium Instruction) will be based on analysis of their academic and professional needs. The methodology covers much ground concerning use of open data as a mean of education process within teaching a language and a subject. This reason led to questions about language awareness alongside use of a non-mother tongue searching data. In this case, English language as well as Russian can be taken as mother tongue (L1) and non-mother tongue (L2).

Although as it is mentioned above open data affects a great variety of areas and in the current situation not only IT specialists have to know how to work with open data and know what it means, but non-IT as well. Thus, non-IT respondents are also in great value, and the further study is aimed to solve their needs and to develop an approach which let improve "digital literacy".

In our research, we asked people about an open data specialist and his or her competences, and two third respondents were puzzled concerning this occupation:

"I have looked at the questionnaire, sent me via email, but I haven't filled it in, as I don't know what the Open Data specialist does. As a variant, I will look it up".

This question was asked by different types of specialists: IT, economy, business, education, banking, etc., however all of them know what open data means. Our respondents were confused mainly by the word "specialist" arguing that there is no such occupation.

Looking through recruiting sites we cannot find a vacancy for open data specialists with exact value in the search whereas there are a great number of positions that are related to open data such as:

Information Service Officer, Senior Grants and Contracts Analyst, Lead Evaluation Expert, Data and Information Management Consultant, Technical Advisor, Research Manager, and so on. So, the ability and skills to work with open data are needed for specialists working in different areas. they can be call professional competences. Although typically and mainly data is related to IT and Computer Science fields and we describe this issue in one of the chapters. Fulfilling research, we found the common requirements such as [2]:

- Coordinate and maintain documentation for award compliance such as environmental,
- Institutional Review Board, Open Data Policy,
- Meta-evaluations utilizing open data source integration,
- Open data and analytics;
- Support data collection and analysis such as Open Data Kit,
- Generate lead thinking and support our transparency and accountability agenda with a specific focus on the role of open data and technology in the fight against corruption.

All above mentioned is a set of competences needed for efficient performance in analytics, sociology, data-journalism as well as IT and CS.

### 2.2 Presentation and Analysis of Data

In this section, we demonstrate the results of the survey on Open Data Specialist competencies. It is arranged due to the questions asked. So, the first is results of the demographic part, the second is devoted to competences the respondents chose as essential and the third illustrates the attitude towards English language as a professional tool.

**Demographic data of respondents.** The following figures present the results on education, age, professional activity and experience of Open Data application.

According to Fig. 1, most of the respondents have got higher education, that is 76,5%, whereas 16,2 % are doctorates.

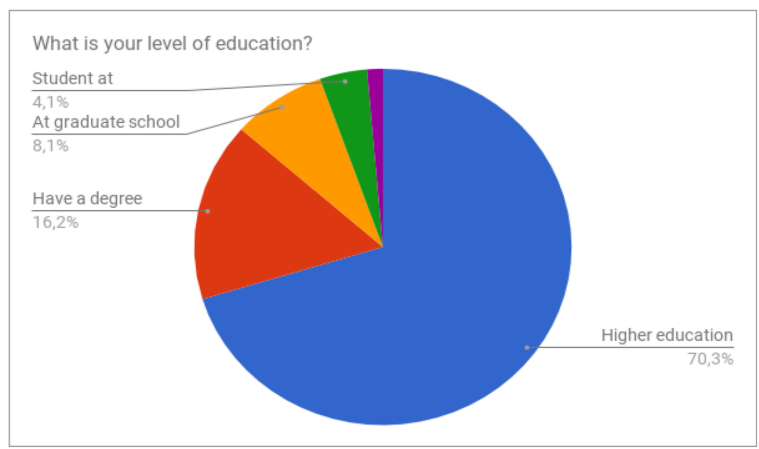

**Fig. 1.** The results of survey about competencies of Open Data specialist. Demographic data. Level of education.

The Fig. 2 provides the results of the survey about respondents age. The respondents are mainly (86,5%) people of active age 18 - 45 years old, whereas the maximum index (39,2%) is 28-39. Although we can see respondents aged over 61 years old.

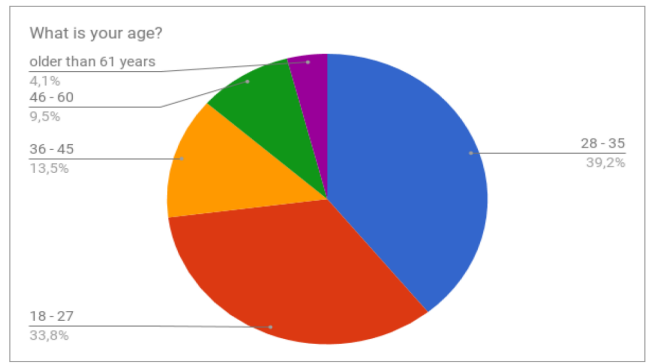

**Fig. 2.** The results of survey about competencies of Open Data specialist. Demographic data. Ages.

Alongside the demographic data, the important matter is a professional field. The Fig. 3 shows the results of the survey concerning professional occupation. The most part of the participants figured out that their professional activity is connected to IT area. Taking into account the channels we used it presents no surprise.

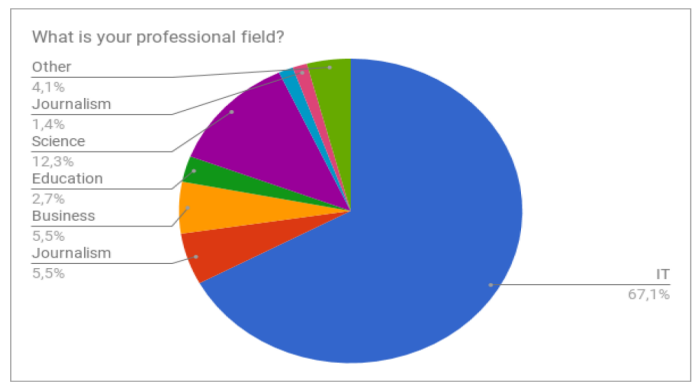

**Fig. 3.** The results of survey about competencies of Open Data specialist. Demographic data. Professional fields.

The last figure in the demographic part presents the results of survey about respondents experience with open data. 87.7 % of them had a clear understanding of what open data is and had experience of its application, meanwhile 20,5% of respondents used it in a professional area. The details are displayed in the Fig.4.

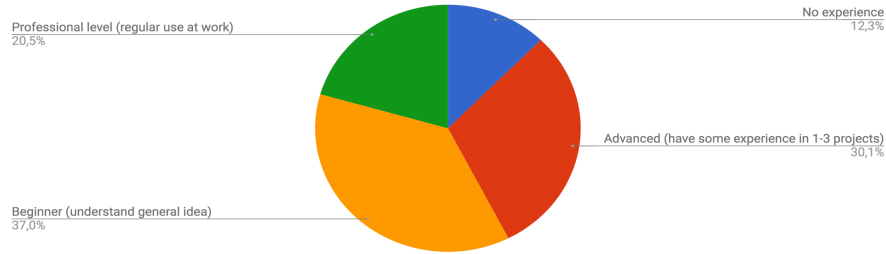

**Fig. 4.** The results of survey about competencies of Open Data specialist. Demographic data. Experience with open data.

**Professional competencies of Open Data Specialist**. The main competences of Open Data Specialist can be clearly seen from the Fig.5.

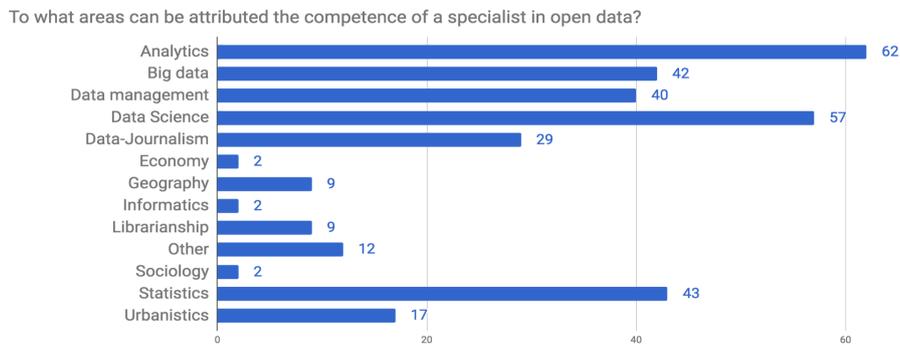

**Fig. 5.** The results of survey about competencies of Open Data specialist. The areas of Open Data specialist competence.

The following table presents the overall results we gained as we conducted the survey. It can have practical value for educator, students and potential employers.

We emphasized the major opinion with bold typeface (see Table 1).

**Table 1.** What competencies should an open data specialist have?

| Competencies and skills | No | Beginner | Advanced |
|---|---|---|---|
| Be able to work with relational DBMSs | 10 | **40** | 24 |
| Be able to work with search engine systems | 0 | 16 | **58** |
| Be able to work with open data platforms | 1 | 18 | **55** |
| Know the basics of statistics | 2 | **41** | 31 |
| Know and be able to use open licenses | 12 | 30 | **32** |
| Be able to program in algorithmic languages | 13 | **36** | 25 |
| Know the SQL language | 10 | **42** | 23 |
| Be able to work with API | 1 | 32 | **41** |
| Be able to work with data processing software | 3 | 32 | **39** |
| Be able to work with libraries of programming languages for data processing | 5 | **35** | 34 |
| Be able to visualize the data | 3 | 31 | 40 |
| Be able to archive the data | 16 | 38 | 20 |
| Be able to work with project management systems | 17 | 44 | 13 |
| Be able to work with version control systems | 13 | 41 | 20 |
| Speak English | 1 | 31 | 42 |
| Be able to work with public data | 6 | 37 | 31 |
| Be able to write HTML parsers | 9 | 36 | 29 |
| Know how to mine the data | 6 | 31 | 37 |

According to Table 1, the most important skills for an open data manager are working with search engines (58%) and open data platforms (55%). Language awareness should be high as an internet research depends on it (42%). Skills of data visualization are important (40%) as well as use a specific API (41%). ODS should be advanced in all above-mentioned skills. Although some specified skills can remain basic. They include working with relational DBMS (40%), understanding statistics (41%) and open licenses (32%). The desirable skill is writing simple queries in SQL (42%) and programs in some algorithmic languages (36%). Open data manager should be a programmer to a certain extent as working with data involves using specific libraries (35%), version control (41%), and project management systems (44%). Knowledge how to parse web pages takes not the least importance (36%) and obtain data from public websites (37%).

**English as a tool in a professional activity.** The question concerning English knowledge as an important skill showed that 89% of respondents found it necessary and use it in their professional field including communication with colleagues.

Although the respondents, asked in our survey, were mainly Russians and we can see that the priority is both Russian and English 61.3% (see Fig. 6), however the search language varies a lot. With respect to results the index of English language, in this case L2, is 26.9%, that is 6 times bigger Russian language, that is L1 for most respondents. Thus, as a search language is used mostly.

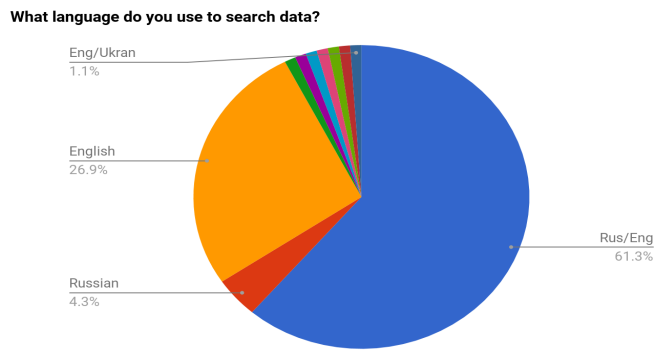

**Fig. 6.** The results of survey about language they use to search data.

The opinion of specialists concerning effectiveness of search language has a particular value for us as it reflects their readiness to gain a new skill. It is clear from the information presented (see Fig.7) English is regarded the most efficient. Whereas the index Depends on Task is in major as well. Fulfilling the task, it is needed to look for data in different resources, and, as one of the respondents noticed, Data is almost never translated. So, in the frame of education this concept of using authentical resources to study L2 in professional sphere is considered the most beneficial. Overall, we need to notice a good sense of humor some experts possess naming SQL.

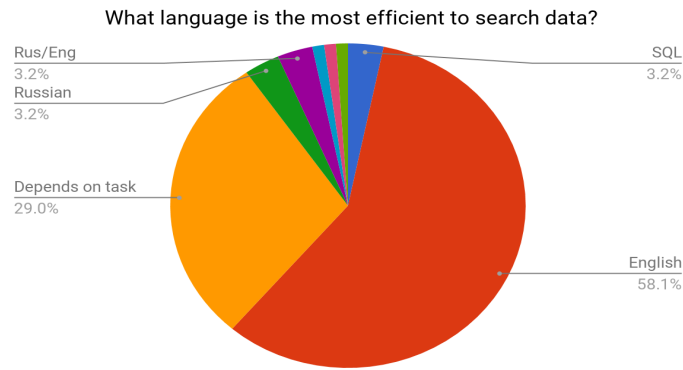

**Fig. 7.** The results of the questionnaire of specialists in different spheres concerning effectiveness of search language.

## 2.3 Discussion

The presented information is a result of hot discussion and can provide us with a deep vision on needed competences of specialists themselves:

- Have to define the limits of data applicability,
- Have good knowledge in a particular field,
- Work with government officials who are responsible for opening data,
- Have to know how to find primary data sources,
- Have some basic knowledge about law,
- Know how to order the information,
- Know how to use regular expressions language.

Also, respondents suggested their idea how to improve this questionnaire. We demonstrate some of interesting suggestions here in Table 2.

**Table 2.** The examples of the suggestions for the survey improving.

| Suggestions |
|---|
| Indicate the name of the specialty received at the university |
| What programming language is relevant to the task of creating open-source databases? |
| The technology stack that was used |
| How do you search for public data? |
| What are your priorities for using some types of open data in practice? |
| What are your priorities for using certain open data models in practice? |
| What software did you use to search and process open data? |
| If possible, indicate the areas of activity in which you carried out projects using open data. |
| Add Gender; The place where the person works. |
| In the table with competences you can add "Availability of DevOps skills" |
| Main areas of application of open data? |
| What kind of programming language, first of all, is it necessary for a specialist in big data |
| The role of the specialist in open data in production chains. |
| Type of education (technical / humanitarian ...) |
| Additional Education (courses / second higher education ...) |
| What is open data? Give an example |

## 3      Job analysis in the sphere of Data and Open Data

To develop a high-class specialist, it is important to know what qualities and skills are in demand and what potential employees have an eye for. We analyzed the most popular recruiting sites such as: indeed.com and LinkedIn. A few statistics was given from CareerBuilder.com, Dice.com and UpWork.com as well.

The information provided in this section below gave a deep understanding of the skill set, dynamics of demand in job market and the salary range.

Indeed.com is job search engine, "currently available in over 60 countries and 28 languages"[1]. It presents the biggest sampling frame.

It reaches 163 place in the rating list in traffic statistics among other Internet recruiting sites due to Alexa, July 2017[2].

### 3.1     The USA search

The number of vacancies searched by keywords features that "Data Manager" is top requested regarding all resources. Meanwhile the difference between "Data Manager" and "Open Data Manager" ranges from 6 to 13. Anyway "Open Data Manager" is still less in-demand. The lowest interest presents "Data Scientist". Table 3 illustrates this to the full extent, as for the salaries, the information provided is not sufficient. Table 4 shows salaries of different specialists in Data Science. We emphasized the largest results and salary with bold typeface.

**Table 3.** Results of vacancies search.

| **Keywords** | **Indeed** | **LinkedIn** | **Career-Builder** | **Dice** | **UpWork** |
|---|---|---|---|---|---|
| **Data Specialist** | 69 000 | 1,670 | **2500+** | 31,706 | 439 |
| **Data Manager** | **206 082** | 1,867 | **2500+** | **40,211** | **3,364** |
| **Open Data Manager** | 34 064 | 188 | **2500+** | 40,162 | 253 |
| **Data Scientist** | 22 899 | **10,968** | 1072 | 30,085 | 158 |

---

[1] https://en.wikipedia.org/wiki/Indeed.com
[2] http://www.alexa.com/siteinfo/indeed.com

Table 4. Salaries (the USA, per year).

| Keywords | Indeed |
|---|---|
| **Data Specialist** | $60,497 |
| **Data Manager** | $95,672 |
| **Open Data Manager** | $95,672 |
| **Data Scientist** | **$130,268** |

### 3.2 The UK search

The given results of job offer and CVs illustrate the same tendency in the UK as in the US, "Open Data Manager" alongside with "Open Data Specialist" reach the indexes in 3 832 and 1 458 respectively, comparing to "Data Manager" and "Data Specialist" which positions are much higher, 37 576 and 14 237 relatively. We emphasized the largest result and salary with bold typeface.

Table 5. Results of vacancies search.

| Keyword for vacancies | Number | Average salary per year |
|---|---|---|
| **Open Data Specialist** | 1 458 | no data about average salary |
| **Data Specialist** | 14 237 | £40,657 |
| **Data Manager** | **37 576** | **£42 405** |
| **Open Data Manager** | 3 832 | no data about average salary |

Table 6. Results of CV search.

| Keyword for CV | Number |
|---|---|
| **Open Data Specialist** | 7 |
| **Data Specialist** | 5506 |
| **Data Manager** | **28063** |
| **Open Data Manager** | 33 |

Key skills for data managers include logic, analytical and problem-solving skills; ability to conduct operations and systems analysis; familiarity with mainframe computers, hard disk arrays and drives; and knowledge of and ability to use archival component-oriented development, operating system, database management and metadata management software programs. [3]

Nowadays this skill set is nothing outrageous but essential for a professional regardless the area he or she works. Unfortunately, this knowledge is typically taught to IT students whereas non-IT specialists experience hardships as soon as they graduate. Feeling the need, they attend courses to develop their competencies.

Getting into this career requires a minimum of a bachelor's degree, but some employers seek individuals with Master of Business Administration degrees. Fields of study include most computer-related fields, such as information technology or computer science. While certification is not required, industry certifications are available. Some employers may require that managers be certified in the software program the organization uses. The experience ranges from 1-5 years of experience in data analysis or database development. [3]

## 4      Open Data is a teaching and learning tool

The growing demand for qualified specialists in such areas as Information Technology and Computer Science make it necessary to monitor the latest developments, participate in international projects, and also constantly improve in the main and related fields. Specialists are forced to "arm" with new knowledge and skills, share information and process data quickly, as these two areas of activity are rapidly developing. As a consequence, various requirements for specialists generate new jobs, sometimes they appear faster than programs in institutions.

One of the key skills of such specialists is the ability to understand and perceive information in L2 (non-mother tongue). Competency to exchange and process data is a key skill, and at this stage it is important to enhance the international language as a core language. The knowledge and ability to apply it can be a critical factor for a professional development and the main reason for making a decision for the employer.

The results of a questionnaire, section 2.2.3, depict that most the respondents choose English as a main language for a typical search they do and as the most efficient language among others. In this case, we should take into consideration that along with Russian participants we questioned citizens from other countries. Presumably being an International language it provides the bigger range of information. Thus, it consequently should be considered as one of the most applicable skill in professional activity which is taught alongside the main content subjects.

Nowadays there is a tendency among educational institutions to teach the basic disciplines in an international language, trying as closely as possible to approximate and simulate the situation of the professional environment. This approach is called EMI (English as Medium Instruction), furthermore we going to have a pilot course at ITMO university using L2 (non-mother tongue).

In this case, several difficulties may arise, first of all, insufficient language competence, and even language awareness does not always lead to the expected result without the possession of a proper methodology. There are several techniques and techniques designed to facilitate the educational process.

Great majority of Institutes relies on ESP (English for Specific Purposes) program using professional literature. Unfortunately, students' books become obsolete too fast to use them for a long period.

Using Open Data as a tool for teaching IT and CS specialists is a solution. As we can see from the survey our respondents use different languages L1 (mother tongue) and L2 (non-mother tongue).

Currently we are working on a framework aimed to help educators both English teachers and subject teachers. In our further work, we will apply Open Data as a tool to teach and enhance both professional and English skills to achieve the greatest result. Talking about methodology we pay an attention to such approaches as Data Expedition, Flipped lessons, Project Oriented Lessons and so on.

Data Expedition is a method, that can bring several benefits to an educational process as, first of all, it gives an opportunity to search information, process and analyze it. As a result, students gain knowledge and experience of work with Open Data, that is an important skill for a qualified specialist regardless of the professional area. Moreover, it allows educators and students to work with up to date Data, which occurs important in rapidly developing world. In fact, the method is very flexible and allows to study either at home or in classroom. In this case, we should pay an attention to various directions such as peer-learning, project based learning etc.

Taking into account difficulties concerning its implementation we need to be sure that an educator is aware of blending learning. The responsibilities related to the content of a lesson or a course becomes more multiple and diverse starting from presenting glossary and finishing with proper technical environment. Overall, this approach needed to be analyzed in the frame of teaching ESP.

In order to obtain a deeper understanding of skills employers wish to see their specialists and our current student have, so the primary competences are the following, we have analyzed some recruiting sites [3] let us take for instance the "Information Service Officer, Bonn, Germany [2]:

**'Must haves': Any professional office, business, commercial, IT related education (e.g. office-, trade-, merchant apprenticeship IHK) or equivalent (university) education (e.g. B.A.) in a relevant subject.**

- Experience with agile development methods (e.g. Scrum),
- Understanding of data modeling, data structure, and data integration,
- Experience with presenting information in reporting/dashboards, BI Tools,
- Strong analytical and conceptual skills,
- Solid coordination and collaboration skills,
- Problem solving attitude,
- Fluency in English written and spoken.

**'Nice to have':**

- Understanding of impact/performance measurement methods,
- Understanding of data quality and quality assurance,
- Experience with Cloud computing, open data sources, and SQL Databases,
- Experience with data integration,
- Good knowledge of any additional language especially Spanish, French, German, Portuguese and/or Italian.

Hardly ever you can find a high qualified specialist job description without an obligatory skill 'good" English alongside with professional requirements. Thus, the list of competences should be wide and studied carefully.

## 5     Discussion

The results of the survey presented in section 2 were hotly discussed in the Telegram channels specialized on Data: OpenDataRussiaChat and Data Science Chat.

The colleagues had questions on the topic of who can be called open data specialist and whether open data specialist should be pointed out as a separate profession. We received some counter proposals allocated to competences in the field of open data.

Many people in the Data Science community do not understand the difference between working with open data and working with data in general.

Example of dialogue in #career channel (in opendatascience Slack) demonstrates this statement:

Is there a distinction and why: "To be able to work with data processing packages" and "To be able to work with libraries of programming languages for data processing"

What kind of skill does it require: "To be able to archive data"?

What is "Know how to work with related public data"?

What for to ask about "To be able to work with relational DBMS" and then more about "To know SQL language"?

## 6     Further works

This study opens our further works in different directions. First of all, we intend to carry out the survey concerning Open Data. Alongside one of our strong goals is to create several courses on Big Data, Data Management, Data Visualization. Secondly, we plan to study more precisely the role of Open Data in ESP and EMI as we aim to develop a framework as a part of pedagogical methodology.

## 7     Conclusion

This study identified the following problems:

- The lack of motivation of the professional community in completing the questionnaire,

- The demand for specialists in OD is still low, but you can track the trend of increasing demand and wages for professionals who are engaged in data analysis, and not just data management.

We found out that there are some jobs straightly connected to Open Data, such as an Open Data Manager which is less popular rather than a Data Manager.

Nevertheless, using Open Data is mostly knowledge which is required to different specialists, being a fundamental skill. Unfortunately, this skill is mainly considered necessary for IT and CS students creating a big gap in "digital literacy" in education of non-IT students.

Whereas we can clearly reveal a portrait of an Open Data Specialist. There is no direct connection to IT sphere, meanwhile this person should be digitally literate, possess logic and the ability to use and work with open data as it affects many areas: sociology, urban studies, geography, statistics, public administration, data journalism, etc. It is especially important to develop and implement training courses on open data for non-IT student. And we believe that it should be accession L1 and L2.

In this paper, we also pay attention to the fact that open data plays a great role as an educational resource, especially in the frame of teaching computer science and Information technology as these fields are constantly developing. Moreover, regarding our preliminary survey we can conclude that open data is mainly searched in English as it is the simplest way to gain necessary information up to date. We face a need to teach CS and IT specialists using both. Open Data can be applied to enhance professional competences as well as to improve English language skills.